\documentclass[aps,prd,nofootinbib,preprintnumbers,twocolumn,showpacs, floatfix]{revtex4-1}
\usepackage{amsmath}
\usepackage{amssymb}
\usepackage{epsfig}
\usepackage{graphicx}
\usepackage{multirow}
\usepackage{color}
\usepackage[normal]{subfigure}
\usepackage{rotating}
\usepackage{hyperref}

\newcommand{\bitem}{\begin{itemize}}
\newcommand{\eitem}{\end{itemize}}
\newcommand{\bwt}{\begin{widetext}}
\newcommand{\ewt}{\end{widetext}}
\newcommand{\be}{\begin{equation}}
\newcommand{\ee}{\end{equation}}
\newcommand{\bdm}{\begin{displaymath}}
\newcommand{\edm}{\end{displaymath}}
\newcommand{\bea}{\begin{eqnarray}}
\newcommand{\eea}{\end{eqnarray}}
\newcommand{\ds}{\displaystyle}
\newcommand{\nn}{\nonumber}

\def\eq#1{{Eq.~(\ref{#1})}}
\def\eqs#1#2{{Eqs.~(\ref{#1})--(\ref{#2})}}
\def\fig#1{{Fig.~\ref{#1}}}

\def\Table#1{{Table~\ref{#1}}}

\def\vev#1{\left\langle #1 \right\rangle}
\def\bra#1{\left\langle #1 \right|}
\def\ket#1{\left| #1 \right\rangle}
\def\qq{\vev{\bar q q}}

\def\Tr{\!\mathop{\rm Tr}}

\def\G{\ensuremath{\bar G}}
\def\M{\ensuremath{\mathcal{M}}}
\def\xQM{$\chi$QM\,}

\begin{document}
\jot = 1.2ex         

\preprint{}
\title{$K\to \pi\pi$ hadronic matrix elements of left-right current-current operators}
\author{Stefano Bertolini}\email{stefano.bertolini@sissa.it}
\affiliation{INFN, Sezione di Trieste, SISSA,
via Bonomea 265, 34136 Trieste, Italy}
\author{Alessio Maiezza}\email{alessio.maiezza@aquila.infn.it}
\affiliation{Universit\`a dell'Aquila, Via Vetoio, Coppito, 67100 L'Aquila, Italy}
\author{Fabrizio Nesti}\email{nesti@aquila.infn.it}
\affiliation{Gran Sasso Science Institute, Viale Crispi 7, 67100 L'Aquila, Italy}
\begin{abstract}
\noindent
Effective $\Delta S=1$ four fermion operators involving left- and right-handed currents are relevant
in left-right gauge extensions of the standard model and scalar extension of the Yukawa sector.
They induce $K\to\pi\pi$ decays which are strictly constrained by experimental data, typically
resulting in strong bounds on the new physics scales or parameters.  We evaluate the $K\to \pi \pi$
hadronic matrix elements of such operators within the phenomenological framework of the Chiral Quark
Model.  The results are consistent with the estimates used in a previous work on TeV scale
left-right symmetry, thus confirming the
conclusions obtained there.\\[-.5ex]
\end{abstract}

\maketitle

\section{Introduction}

\noindent
The enduring successes of the Standard Model (SM), in particular in the quark flavor sector,
naturally provide stringent tests and constraints on New Physics (NP) theoretical modeling.  An
historical and relevant role in testing NP models is played by the CP-violating $\Delta S=1,2$
processes in Kaon physics.  It is in fact a common feature of NP scenarios the presence of
additional flavor-changing (FC) interactions that induce new effective operators at the electroweak
scale.  The presence of such operators often leads to sharp constraints on the scales and the
couplings of the extended theory.

A paradigmatic example is the Left-Right (LR) extension of the SM~\cite{Pati:1974yy,
  Mohapatra:1974hk, Mohapatra:1974gc, Senjanovic:1975rk, Senjanovic:1978ev}, which in its minimal
version provides a complete theory of neutrino masses~\cite{Nemevsek:2012iq}, and directly connects
possible new accelerator (LHC) physics to lower energy phenomena like neutrinoless double-beta decay
and lepton flavor violation~\cite{Tello:2010am} (for a recent review see e.g.\
Ref.~\cite{Senjanovic:2011zz}). In the quark sector, flavor changing operators lead to a lower bound
on the right-handed gauge boson scale slightly above the TeV region~\cite{Maiezza:2010ic}, within
the reach of LHC searches. LR symmetric models generate a new set of FC operators (a complete set
for $\Delta S=1$ can be found in~\cite{Bertolini:2012pu}) some of which turn out to be crucially
relevant for phenomenology.

Characteristic of the LR setup are the following current-current operators,
\bea
Q_1^{LR} &=& (\bar{s}_\alpha u_\beta)_L (\bar{u}_\beta d_\alpha)_R \quad\;  Q_1^{RL} = (\bar{s}_\alpha u_\beta)_R (\bar{u}_\beta d_\alpha)_L  \nn \\
Q_2^{LR} &=& (\bar{s} u)_L (\bar{u} d)_R \qquad\quad\ \, Q_2^{RL} = (\bar{s} u)_R (\bar{u} d)_L\,,
\label{Q1Q2LR}
\eea
where the subscripts $L,R$ stand for $\gamma_\mu (1\pm \gamma_5)$ and $\alpha,\beta$ are color
indices, understood in $Q_2$. In the LR models these operators are generated at tree level by gauge
boson exchange, and thus have a prominent role in setting constraints on the model
parameters~\cite{Ecker:1985vv, Maiezza:2010ic}.  They are as well generated in other popular NP
extensions of the SM, as for instance supersymmetry (SUSY), with FC processes driven by squark
mediation, or extended Higgs models with FC interactions (for a systematic discussion on flavor
physics beyond the SM see Ref.~\cite{Buras:2011}).

As mentioned above the operators (\ref{Q1Q2LR})\ play a role in $\Delta S=1$ processes and they are
particularly relevant for the study, within NP models, of direct CP violation in $K\to \pi\pi$
decay, namely for the calculation of the $\varepsilon'$ parameter. In order to match the experimental
precision, the $K\to \pi\pi$ matrix elements of the effective operators are needed beyond the simple
and naive factorization.  First principle approaches to non-perturbative QCD (of which lattice is
the foremost) have not yet provided an accurate and reliable answer. In this work we address this
issue by offering a (phenomenological) calculation of such a matrix elements in the framework of the
Chiral Quark Model (\xQM)~\cite{Nishijima:1959uu, Gursey:1959yy, Gursey:1961uu, Cronin:1967jq,
  Weinberg:1978kz, Manohar:1983md, Manohar:1984uq, Espriu:1989ff}.

To this aim, we construct and determine, via the integration of the constituent quark fields of the
\xQM, the $\Delta S=1$ chiral lagrangian relevant to $Q_{1,2}^{LR,RL}$ at $O(p^2)$ in the momentum
expansion. The chiral coefficients are generally determined within the model in terms of three non
perturbative parameters, namely the constituent quark mass $M$, the quark condensate $\qq$ and the
gluon condensate $\vev{\tfrac{\alpha_s}{\pi}GG}$. Their values and model ranges were
phenomenologically determined in Ref.~\cite{Bertolini:1997ir} via the fit of the $\Delta I=1/2$
selection rule in $K\to \pi\pi$ decays, what eventually lead to the successful prediction of
$\varepsilon'/\varepsilon$~\cite{Bertolini:1997nf}. Such a phenomenological and self-contained
determination of the model parameters represents in our opinion the strength of the approach and it
is at the root of the robustness of the results.

\pagebreak[3]

Consistency with the needed order in momentum expansion requires the inclusion of the chiral loops
contributions to the $K\to\pi\pi$ amplitudes, that we compute. Eventually, we provide the $O(p^2)$
matrix elements via the $B$-parameters, which gauge the departure from the Vacuum Saturation
Approximation (VSA).  The $B$-parameters are given at the intrinsic \xQM scale of about $0.8\,$GeV,
as well as at 2\,GeV, for direct comparison with forthcoming lattice calculations.

\section{The $\Delta S=1$ chiral lagrangian (making of)}

\noindent
The quark $\Delta S=1$ effective lagrangian is written as a combination of local quark operators
\begin{equation}
\mathcal{L}_{\Delta S=1} =\Sigma_{i} C_i (\mu)Q_i (\mu) \,,
\label{LDS1eff}
\end{equation}
where $Q_i$ are effective four-quark operators as in \eq{Q1Q2LR} and $C_i$ are their short-distance
Wilson coefficients, evaluated at a scale $\mu$. By extending the SM to include RH interactions, the
sum in \eq{LDS1eff} spans a complete set of twenty-eight operators~\cite{Bertolini:2012pu}, which
exhibit all chiral combinations of $L,R$ currents.

Contact with the physical mesonic transitions is made once the relevant hadronic matrix elements of
the effective quark operators are computed. Since the relevant scale for kaon physics falls in the
strong interacting regime of QCD, the problem cannot be addressed with perturbative (coupling
expansion) methods.  In the present work we address this issue by means of a phenomenological
approach based on the \xQM.

The \xQM takes advantage of the QCD chiral symmetry while introducing an effective quark-meson
interaction. This provides a bridge between the perturbative QCD and chiral lagrangian regimes.  The
model can be seen as the mean-field approximation of an extended Nambu-Jona-Lasinio model that
mimics QCD at intermediate energies~\cite{Bijnens:1992uz,Bijnens:1995ww}.  After integrating out the
constituent quark fields, the meson octet interactions are determined in terms of three non
perturbative parameters: the constituent quark mass, the quark condensate and the gluon condensate.
The model is renormalizable in the large-$N_c$ limit~\cite{Weinberg:2010bq} and it is successful in
reproducing the $O(p^4)$ low energy constants of the Gasser-Leutwyler lagrangian as well as a number
of observables, albeit one must be aware of its limitations~\cite{deRafael:2011ga,Greynat:2012ww}.

In the nineties a thorough investigation of the $\Delta S=1$ and $\Delta S=2$ chiral lagrangians
within the framework of the \xQM has been carried out~\cite{Antonelli:1995nv, Antonelli:1996qd,
  Bertolini:1997ir, Bertolini:1997nf}. The project led to a successful prediction of the direct CP
violation in $K\to\pi\pi$ decays ($\varepsilon '/\varepsilon$) shortly before its experimental
determination. The approach was based on the self-consistent determination of the non perturbative
parameters of the \xQM via the fit of the CP conserving $\Delta I=1/2$ selection rule in
$K\to\pi\pi$.  Such a phenomenological setup was central to reducing the model systematic
uncertainties and to providing a robust prediction.

While, ultimately, first principle approaches to non perturbative QCD (lattice being the foremost,
see Ref.~\cite{Boyle:2012ys} for recent developments) must provide the evaluations of hadronic
transitions, here we apply the \xQM phenomenological approach to the calculation of
the $K\to\pi\pi$ matrix elements of the LR current-current operators (\ref{Q1Q2LR}).

\subsection{The Chiral Quark Model}

\noindent
In the \xQM\ a meson-quark interaction term is added to the ordinary QCD
lagrangian:
\begin{equation}\label{mesonquark}
\mathcal{L}_M=-M\big(\bar{q}_R\Sigma q_L+\bar{q}_L\Sigma^\dag q_R\big)\,,
\end{equation}
where $q=(u\ d\ s)^t$ and $\Sigma\equiv e^{\frac{2i}{f}\Pi(x)}$, $\Pi(x)$ being the $SU(3)$ meson
octet acting on the fundamental representation.  The parameter $M$ is identified as the constituent
quark mass, as it follows from a chiral quark rotation that absorbs $\Sigma$ in the constituent
quark fields (henceforth referred to as the ``rotated" picture):
\begin{equation}\label{puremass}
\mathcal{L}_M=-M\big(\overline{Q}_R Q_L+\overline{Q}_L Q_R\big)\,,
\end{equation}
where $q_{L}=\xi^\dag Q_{L}$, $q_{R}=\xi Q_{R}$, with $\Sigma=\xi^2$ and $\Sigma^\dag=(\xi^\dag)^2$
respectively.  In the rotated picture the quark-meson interactions arise from the quarks kinetic term,
as
\begin{equation}\label{int}
\mathcal{L}_{int}=\overline{Q}(\gamma^\mu V_\mu+\gamma^\mu \gamma_5 A_\mu)Q\,,
\end{equation}
the vector and axial fields $V_\mu$ and $A_\mu$ being defined as
\begin{eqnarray}
V_\mu &=& \frac i2 (\xi^\dag \partial_\mu \xi +\xi \partial_\mu \xi^\dag) \nn \\
A_\mu &=& \frac i2 (\xi^\dag \partial_\mu \xi-\xi \partial_\mu \xi^\dag)\,.
\label{Amu}
\end{eqnarray}

Analogously, in the rotated description the ordinary quark mass term
\begin{equation}
\mathcal{L}_m=\bar{q}_R \M q_L+\bar{q}_L \M^\dag q_R\,,
\end{equation}
with $\M={\rm diag}\{m_u,m_d,m_s\}$, becomes
\begin{equation}\label{mass}
\mathcal{L}_m=\overline{Q}_R  \xi^\dag \M \xi^\dag   Q_L+\overline{Q}_L  \xi \M^\dag \xi  Q_R\, .
\end{equation}

\subsection{Bosonic representation of the quark operators}

\noindent
The \xQM provides a systematic way of constructing the bosonic representation of the effective
$\Delta S =1$ quark operators in \eq{LDS1eff}~\cite{Antonelli:1995nv}.  By integrating out the
constituent quarks, an effective chiral lagrangian is generated
\begin{equation}\label{mesonlevel}
\mathcal{L}_{\Delta S=1}=\sum_{i,j}G_j(Q_i)O_j^{\chi}\,,
\end{equation}
where $O_j^{\chi}$ are bosonic operators involving the octet meson fields and $G_j(Q_i)$ are the chiral
coefficients determined by the matching with the \xQM lagrangian.

We are now set to construct the chiral representation of the $Q_{1,2}^{LR}$ operators. It is
sufficient to consider the LR operators, as the RL ones are related to the former by
symmetry.  Up to the color structure, both $Q_{1,2}^{LR}$ have the form
\begin{equation}
\label{eq:unrotatedQ}
\bar{q}_L \lambda_1^3 \gamma^\mu q_L \,  \bar{q}_R \lambda_2^1 \gamma_\mu q_R
\end{equation}
that in the rotated picture reads
\begin{equation}\label{eq:rotatedQ}
\overline{Q}_L \xi \lambda_1^3 \gamma^\mu  \xi^\dag Q_L \, \overline{Q}_R \xi^\dag \lambda_2^1 \gamma_\mu  \xi Q_R\,.
\end{equation}
The flavor projectors $\lambda_i^j$ are appropriate matrices such that $\bar{q} \lambda_i^j
q=\bar{q}_j q_i$, for $i,j=1,2,3$.

For any such four quark operator, the effective bosonic operators in the chiral lagrangian arise by
integrating out the quarks $Q$, while inserting in all possible ways either two $A_\mu$ fields
of~\eq{Amu}, or $\xi^\dag \M \xi^\dag$, $\xi \M^\dag \xi$ from~\eq{mass} in the constituent quark loops.
Since $V_\mu$ transforms as a gauge field, terms involving the vector field break local chiral
invariance and cannot appear in the $O(p^2)$ chiral lagrangian~\cite{Weinberg:1978kz,
  Manohar:1983md, Manohar:1984uq}. Both the operators in \eq{eq:rotatedQ} and their fierzed forms
can be used~\cite{Antonelli:1995nv}.  By applying this procedure to $Q_{1,2}^{LR}$ we obtain at
$O(p^2)$
\begin{eqnarray}
\ \ &&\!\!\!\!\mathcal{L}_{\Delta S=1}
= \G_0(Q_{1,2}^{LR}) \,\Tr \big[\lambda_1^3 \Sigma^\dag \lambda_2^1 \Sigma \big]\nn \\
&&{}+ \G_m(Q_{1,2}^{LR}) \,\big\{\Tr\! \big[\lambda_2^1 \Sigma \lambda_1^3 \Sigma^\dag  \M \Sigma^\dag \big]\!+\Tr\! \big[\lambda_1^3 \Sigma^\dag \lambda_2^1 \Sigma  \M^\dag \Sigma \big]\!\big\} \nn \\
&&{}+ \G_{LR}^a(Q_{1,2}^{LR}) \,\Tr \big[\lambda_2^3 D^\mu \Sigma \big] \,\Tr \big[\lambda_1^1 D_\mu \Sigma^\dag \big] \nn \\
&&{}+ \G_{LR}^b(Q_{1,2}^{LR}) \,\Tr \big[\lambda_1^3 \Sigma^\dag D^\mu \Sigma \big] \,\Tr \big[\lambda_2^1 \Sigma D_\mu \Sigma^\dag] \nn \\
&&{}+ \G_{LR}^c(Q_{1,2}^{LR}) \,\big\{\Tr \big[\lambda_2^3 \Sigma \big] \,\Tr \big[\lambda_1^1 D_\mu \Sigma^\dag D^\mu \Sigma \Sigma^\dag \big] \nn \\
&&\qquad\qquad\qquad {}+\Tr \big[\lambda_2^3 D_\mu \Sigma D^\mu \Sigma^\dag \Sigma \big] \,\Tr \big[\lambda_1^1 \Sigma^\dag \big]\big\}\,,
\label{l.chiral}
\end{eqnarray}%
where we used $A_\mu=-\frac{i}{2}\xi (D_\mu \Sigma^\dag) \xi= \frac{i}{2} \xi^\dag (D_\mu \Sigma)
\xi^\dag$, and flavor trace rearrangements~\cite{Antonelli:1995nv}.

The term proportional to $\G_0$ corresponds to no axial field insertion. The terms $\G_{LR}^{a,b,c}$
arise from the insertion of two axial fields $A_\mu$, while $\G_m$ corresponds to the insertion of
$\mathcal{L}_m$, \eq{mass}~\cite{Bertolini:1997nf}.  We use the notation $\G$ to distinguish them
from the analogous SM chiral coefficients $G$ in Ref.~\cite{Antonelli:1995nv}. In our calculation we
take $m_u=m_d=0$, so that the relevant contribution to $\G_m$ is proportional to $m_s$.

\section{Calculation of the chiral coefficients}\label{sec:chiralcoef}

\noindent
In the \xQM\ the amplitudes for processes involving external mesons are evaluated through quark
loops connected by a given operator insertion, and quark-meson interactions as given for instance by
\eq{mesonquark} in the unrotated picture.  The contribution to the chiral coefficients $G_i$ of a
given quark operator is computed by matching the \xQM\ amplitude for a conveniently chosen mesonic
transition with the same amplitude obtained from the expansion of the chiral
lagrangian~(\ref{l.chiral}).

At order $O(p^2)$, and for the operators considered, five coefficients $\G_0$, $\G_m$,
$\G^{a,b,c}_{LR}$ are present, thus requiring five independent matching equations. We will choose
below to calculate the off-shell transitions $K^0\to\pi^0$ and $K^+\to\pi^+$, together with the
on-shell $K\to\pi^+\pi^-$, $K\to\pi^0\pi^0$ decay amplitudes. Expanding in the quark mass one of the
off-shell transitions will then determine $\G_m$.

For the regularization of the divergent integrals we use dimensional regularization ($d=4-2 \epsilon
$). The ``quadratic" ($\epsilon-1$ pole) and logarithmic ($\epsilon$ pole) divergences serve
as a bookkeeping device to identify the ``bare'' quark condensate and the meson decay constant,
namely~\cite{Bijnens:1992uz,Bijnens:1995ww}
\bea
\qq^{(0)} &=& -\frac{N_cM^3}{4\pi^2}\left(\frac{4\pi\tilde{\mu}^2}{M^2}\right)^\epsilon \,\Gamma(\epsilon-1)\,,
\label{bareqq}
\\[.4ex]
f^{(0)} &=& \frac{N_cM^2}{4\pi^2f}\left(\frac{4\pi\tilde{\mu}^2}{M^2}\right)^\epsilon \,\Gamma(\epsilon)\,,
\label{baref}
\eea
where $N_c$ is the number of colors. We replace below these bare parameters with the physical ones
$\qq$ and $f$. This automatically includes in the chiral coefficients the factorizable gluon
condensate corrections that are needed to recover the numerical consistency of the parameters, in
particular of $\qq$.
It is also worth mentioning that \eqs{bareqq}{baref} imply a relation between the two bare
quantities. This ambiguity, intrinsic to the regularization scheme, becomes numerically irrelevant
when all the contributions to the amplitudes at a given order are included.
Finally, at the one loop level in the chiral expansion (and at $O(p^2)$ in the momentum expansion)
$f$ will be further identified with the renormalized decay constant $f_1$, that reproduces the
correct pion and kaon decay constants~\cite{Bertolini:1997ir}.

\subsection{Constituent quark loops}
\label{sec:cql}

\begin{figure}
\center
\centerline{\includegraphics[width=.99\columnwidth]{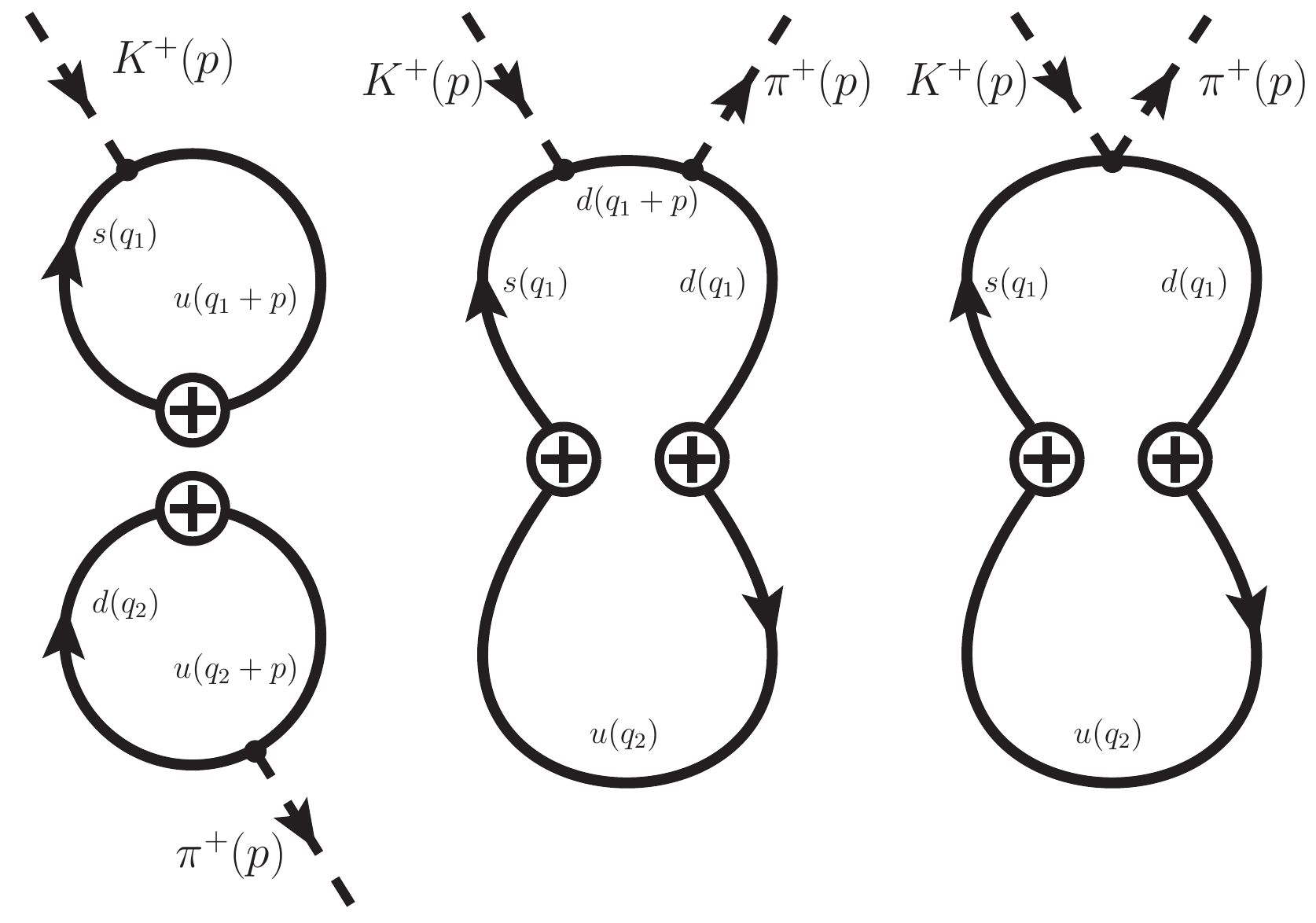}}
\vspace*{-1.7ex}
\caption{Diagrams leading to the off-shell $K^+ \to \pi^+$ transition within the \xQM, in the
  unrotated quark picture. The black points represent the meson-quark vertices, while the crossed
  circles represent the four-quarks operator insertion. The loop momenta are $q_1$ and $q_2$.  The
  two configurations of constituent quark loops correspond to the product of two distinct Dirac
  traces or to a single one.}
\label{Offshelldecay}
\end{figure}

\noindent
With one insertion of the four-fermion operators, the constituent quark loops appear in two possible
patterns, the so called factorized or the unfactorized form~\cite{Antonelli:1995nv}, corresponding
to two distinct or a single Dirac trace. The amplitude is evaluated considering all the possible
ways the desired process is realized by attaching the meson fields to the quark loops. This is best
performed in the unrotated picture.  As an example the off-shell process $k^+ \to \pi^+$ is
represented in \fig{Offshelldecay}.  The way color indices are saturated in the quark operator
determines which diagrams are leading or subleading in $1/N_c$. For the case of $Q_2^{LR}$, the
first diagram in \fig{Offshelldecay} is of $O(N_c)^2$ whereas the last two are of $O(N_c)$. The
opposite occurs for $Q_1^{LR}$.

The direct computation in the Naive Dimensional Regularization (NDR) $\gamma_5$-scheme of the
diagrams in \fig{Offshelldecay} for the $Q_{1,2}^{LR}$ operators leads to
\bea
\langle \pi^+| &Q_1^{LR}& |K^+\rangle_{\rm NDR}=\frac{2i}{3}\left[
3\qq \left(\frac{\qq}{f^2}+m_s\right)- f^2 p^2\right]
   \nn  \\
&&\quad{} -\frac{6i M^2}{\Lambda_\chi ^2} \Big[2 f^2 p^2+M \left(f^2 m_s+\qq\right)\Big]\,,
\label{piQ1kNDR}
\eea
and
\bea
\langle \pi^+| &Q_2^{LR}& |K^+\rangle_{\rm NDR}=\frac{2 i}{3} \left[\qq \left(m_s
+\frac{\qq}{f^2}\right)- 3 f^2 p^2\right]   \nn  \\
&&{}-\frac{2 i M^2}{\Lambda_\chi ^2} \Big[2 f^2 p^2+M \left(f^2 m_s+\qq\right)\Big]\,,
\label{piQ2kNDR}
\eea
where $\Lambda_\chi=2\pi \sqrt{6/N_c} f_\pi \simeq 0.82\,$GeV~\cite{Antonelli:1995nv} is the
natural cutoff of the theory and $f_\pi$ is the pion decay constant. The two amplitudes exhibit a
leading term and a subleading one in the $M^2/\Lambda_\chi^2$ expansion.

Similar results are found in the 't Hooft-Veltman (HV) $\gamma_5$-scheme:
\bea
\langle \pi^+|&Q_1^{LR}&|K^+\rangle_{\rm HV}=\frac{2i}{3}  \left[3 \qq \left(\frac{\qq}{f^2}+m_s\right)- f^2 p^2\right]    \nn  \\
&&\qquad\qquad\qquad{}-\frac{12i M^2 f^2 p^2}{\Lambda_\chi ^2}\,,
\label{piQ1kHV} \\[.7ex]
\langle \pi^+|&Q_2^{LR}&|K^+\rangle_{\rm HV}=\frac{2 i}{3} \left[\qq \left(m_s
+\frac{\qq}{f^2}\right)- 3 f^2 p^2\right]   \nn  \\
&&\qquad\qquad\qquad{}-\frac{4 i M^2 f^2 p^2}{\Lambda_\chi ^2}\,,
\label{piQ2kHV}
\eea
which differ from \eqs{piQ1kNDR}{piQ2kNDR} only by subleading terms in $M^2/\Lambda_\chi ^2$.

The corresponding expressions for the $K\to\pi^+\pi^-$ and $K\to\pi^0\pi^0$ on-shell processes are
also found by direct evaluation of the quark loops in the unrotated picture. These processes involve
a fairly large number of diagrams and we do not report here the detailed expressions.

\begin{table}
$
\begin{array}{lcc}
\hline\hline
\hspace*{.25\columnwidth}   &\hspace*{.11\columnwidth} \rm HV \hspace*{.11\columnwidth} &\rm NDR \\[.4ex]
 \hline \\[-2ex]
\G_0(Q_1^{LR})     &\ds -2 \qq^2                     &\ds -2\qq^2  \left(1-\frac{3 M^3 f^2}{\qq \Lambda_\chi^2}\right)           \\[2.2ex]
\G_m(Q_1^{LR})    &\ds -2 f^2 \qq                    &\ds -2f^2 \qq \left(1-\frac{3 M^2}{\Lambda_\chi^2}\right)         \\[2.2ex]
\G_{LR}^a(Q_1^{LR})&\ds \phantom{-}2\frac{f^2 \qq}{M} &\ds \phantom{-}2\frac{f^2\qq}{M} \left(1-\frac{3 M^2}{\Lambda_\chi^2}\right)         \\[2.2ex]
\G_{LR}^b(Q_1^{LR})&\ds-\frac{f^4}{3}                 &\ds -\frac{f^4}{3}                                            \\[2.2ex]
\G_{LR}^c(Q_1^{LR})&\ds-\frac{6Mf^2\qq}{\Lambda_\chi^2} &\ds -\frac{6Mf^2\qq}{\Lambda_\chi^2}                         \\[3ex]
\G_0(Q_2^{LR})    &\ds  -\frac{2}{3} \qq^2       &\ds -\frac{2}{3} \qq^2   \left(1-\frac{3 M^3 f^2}{\qq \Lambda_\chi^2}\right)    \\[2.2ex]
\G_m(Q_2^{LR})    &\ds  -\frac{2}{3} f^2 \qq    &\ds -\frac{2}{3} f^2 \qq \left(1-\frac{3 M^2}{\Lambda_\chi^2}\right)  \\[2.2ex]
\G_{LR}^a(Q_2^{LR})&\ds  \phantom{-}\frac23 \frac{f^2 \qq}{ M}    &\ds\phantom{-}\frac23\frac{f^2 \qq}{M}  \left (1-\frac{3 M^2}{\Lambda_\chi^2}\right)  \\[2.2ex]
\G_{LR}^b(Q_2^{LR})&\ds  \vphantom{\frac{f^2}M}f^4           &\ds f^4                                           \\[2.2ex]
\G_{LR}^c(Q_2^{LR})&\ds   -\frac{2Mf^2\qq}{\Lambda_\chi^2}   &\ds -\frac{2Mf^2\qq}{\Lambda_\chi^2}               \\[3ex]
\hline\hline
\end{array}
$
\caption{The contributions of $Q_{1,2}^{LR}$ to the chiral coefficients in \eq{l.chiral} as computed 
  in the \xQM, in the HV and NDR renormalization schemes.}
\label{tab:G}
\end{table}

By matching all mesonic amplitudes with the corresponding transitions obtained from the expansion of
the chiral lagrangian~(\ref{l.chiral})\ we determine the chiral coefficients $\G_0$, $\G_m$ and
$\G_{LR}^{a,b,c}$ up to order $O(p^2)$.  The results are reported in \Table{tab:G}, in both the NDR
and HV $\gamma_5$-schemes.

The chiral coefficients depend on the quark condensate $\qq$ and on the $f$ parameter. The latter
will be eventually identified, after inclusion of the $O(p^2)$ wave function renormalization and of
the chiral loop corrections to the LO term ($\G_0$) of the amplitude, with the $O(p^2)$ decay
constant parameter $f_1$~\cite{Bertolini:1997ir}.

Some subtleties arise during the calculation. Namely, in the NDR scheme one is not allowed to use
Fierz rotations and, as consequence, both the factorized and unfactorized calculations have to be
performed. In the HV scheme fierzing is allowed that simplifies the calculation, but one must be
aware of the possible presence of ``fake'' chiral anomalies (see for
instance~\cite{Chanowitz:1979zu}), and convenient subtractions have to be implemented in the chiral
lagrangian~\cite{Antonelli:1995nv}.  This implies, among else, that $\G^b_{LR}$ can be computed in
both schemes from factorized diagrams, and as a consequence it does not depend on the
$\gamma_5$-scheme.  This holds also for $\G^c_{LR}$ that turns out to be subleading in
$M^2/\Lambda_\chi^2$. In this case, the vanishing of the leading contribution is immediately seen by
considering the bosonization of the fierzed operator.

It is interesting to mention that the results for the chiral coefficients induced by the $8_L\otimes
8_R$ $Q_{1,2}^{LR}$ operators are common to any four quark LR operator of the
form~(\ref{eq:unrotatedQ}), with an arbitrary choice of flavor projectors, which transform in
general as $(8+1)_L\otimes(8+1)_R$. This is best understood in the rotated picture, where one can
treat the flavor projectors $\lambda_i^j$ as spurions, that eventually appear in the diverse
operators in the chiral lagrangian~(\ref{l.chiral})\ without affecting the values of the
coefficients. In practice the coefficients are $SU(3)$ invariant, as long as they depend on chirally
symmetric parameters.  One may in fact verify that the results given in \Table{tab:G} for
$Q_{1,2}^{LR}$ coincide with those obtained in Refs.~\cite{Bertolini:1997ir,Bertolini:1997nf} for
the operators $\Delta Q_{8,7}$ that differ in the flavor structure and transform as
$8_L\otimes(8+1)_R$.  This result is nontrivial in the unrotated picture as the number and the
topology of the diagrams involved in the bosonization and in the computation of the chiral
coefficients for the two sets of operators does crucially depend on the flavor indices.  In passing,
let us also mention that when computing the isospin 2 component of the $K\to\pi\pi$ amplitude the
difference between the $Q_{1,2}^{LR}$ and $\Delta Q_{8,7}$ operators vanishes.
In the chiral limit this implies a relation among the $Q_{1,2}^{LR}$ and $Q_{8,7}$ contributions to
$A_2$ on which we will comment in the following.

In \Table{tab:G} subleading $1/N_c$ corrections due to the gluon condensate
$\vev{\tfrac{\alpha_s}{\pi}GG}$ are neglected.  The values of the non-perturbative \xQM parameters,
namely $M$, $\qq$ (and $\vev{GG}$) are phenomenologically determined by the successful fit of the
$\Delta I=1/2$ selection rule in $K\to\pi\pi$, as explained in Ref.~\cite{Bertolini:1997ir}.

\section{$K^0\to\pi\pi$ matrix elements at $O(p^2)$}

\begin{figure}
\center
\includegraphics[scale=0.45]{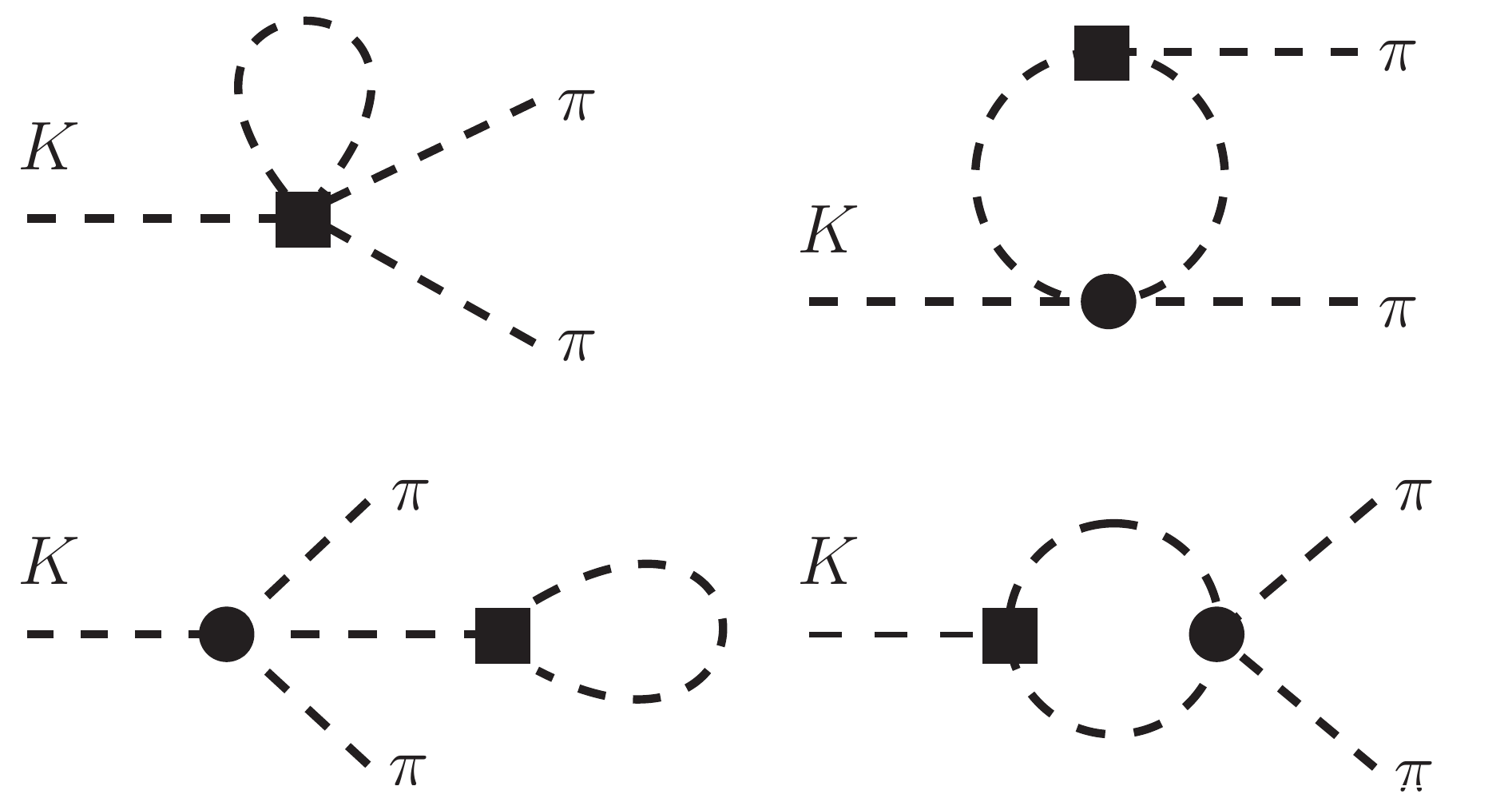}
\caption{Chiral loop vertex renormalization of $K\to\pi\pi$. The internal states are all the allowed
  $SU(3)$ octet mesons. The square box represents the weak vertex, while the circle represents the
  insertion of a strong vertex.}
\label{renvertex}
\end{figure}

\begin{figure}
\center
\includegraphics[scale=.45]{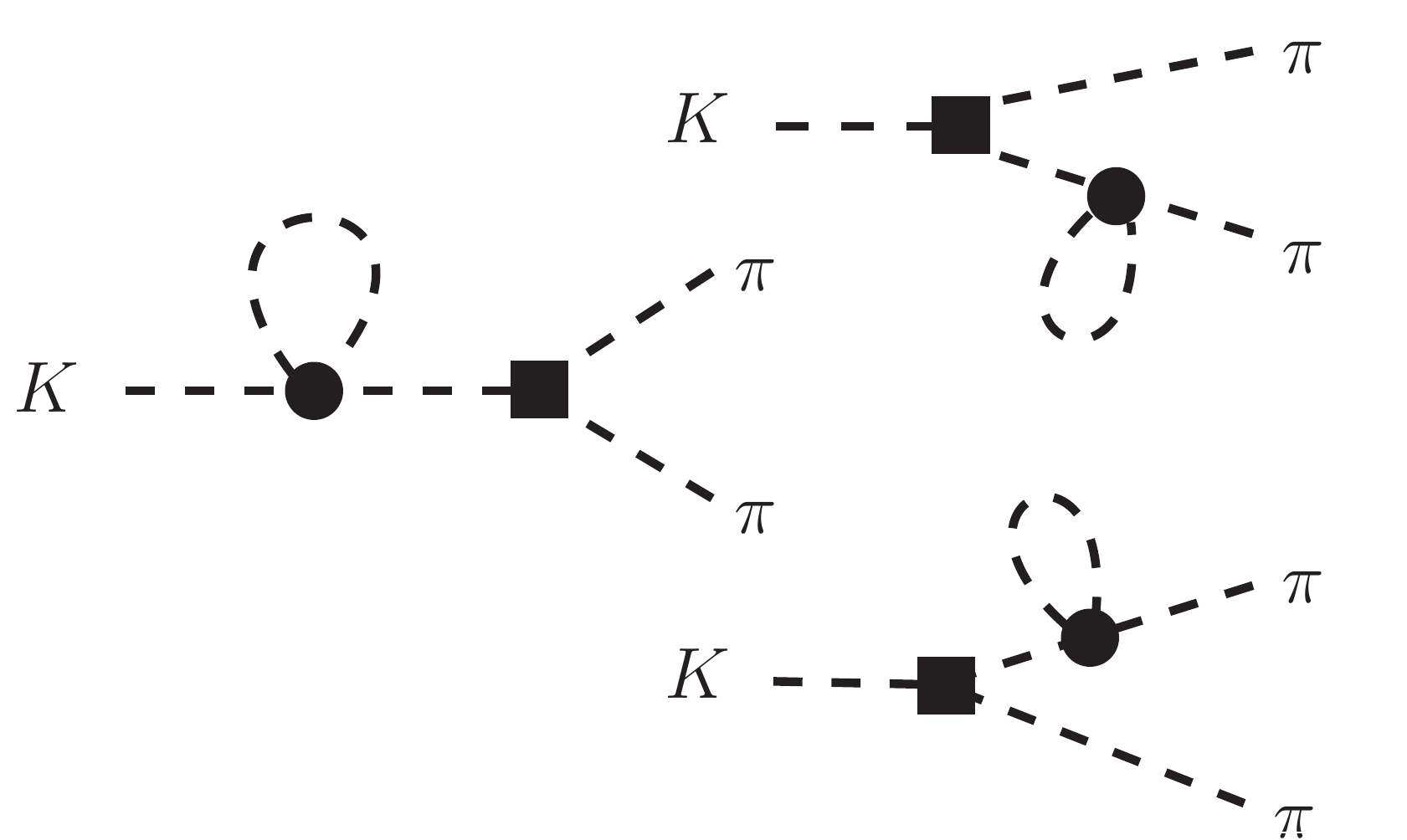}
\caption{Chiral loop wave function renormalization for the $K\to\pi\pi$ transitions. The square box
  and the circle represent the weak and the strong vertex respectively.  All allowed $SU(3)$ octet
  mesons are exchanged in the loop.}
\label{renwf}
\end{figure}

\subsection{Chiral loops}

\noindent
In the previous section we have computed the chiral coefficients, induced by the quark operators
$Q_{1,2}^{LR}$, at $O(p^2)$ which, in the case at hand, is NLO in the momentum expansion.  In order
to consistently compute the $K\to\pi\pi$ amplitudes in the chiral expansion, we must include the
corrections due to the relevant chiral loops. Again, it is enough to focus on the $Q_{1,2}^{LR}$ operators,
since the matrix elements of the RL ones have the opposite sign due to parity.

Using the standard decomposition of the $K\to\pi\pi$ amplitudes in isospin zero and two, $A_0=(2
A_{\pm} +A_{00} )/\sqrt{6}$, $A_2=(A_{\pm}- A_{00} )/\sqrt{3})$, it is also useful to parametrize
them as
\be\label{eq:paramA}
A_{0,2} = A_{0,2}^{\chi_0}+A_{0,2}^{\chi_1}\,,
\ee
where the superscripts $\chi_{0,1}$ refers to the tree and 1-loop chiral contributions respectively.

The tree-level isospin amplitudes for both $Q_{1,2}^{LR}$ read
\bea
\label{A0tree}
A_0^{\chi_0}& =& \frac{1}{\sqrt{3} f^3} \bigg[4 \G_0 Z_\pi \sqrt{Z_K} +\!4 \G_m m_s+\!\G^a_{LR}\! \left(3 m_K^2\!+\!m_\pi^2\right)  \nn\\
&& {}+ 2 \G^b_{LR} \left(m_K^2-m_\pi^2\right)+\G^c_{LR} \left(2 m_K^2-9 m_\pi^2\right)\bigg]
\\[1ex]
\label{A2tree}
A_2^{\chi_0} &=& -\frac{1}{ f^3} \sqrt{\frac{2}{3}}\bigg[ \G_0 Z_\pi \sqrt{Z_K}+\G_m m_s+\G_{LR}^a m_\pi^2 \nn \\
&&\qquad \qquad{} +\G_{LR}^b \left(m_K^2-m_\pi^2\right)-\G_{LR}^c m_K^2\bigg]\,,
\eea
where $Z_\pi$ and $Z_K$ are the one-loop wavefunction renormalizations within the
\xQM~\cite{Bertolini:1997ir}:
\be
Z_\pi = 1-2 \frac{m_\pi^2}{\Lambda_\chi^2}\,,\qquad
Z_K = 1-2 \frac{m_K^2}{\Lambda_\chi^2}+6 \frac{M m_s}{\Lambda^2_\chi}\,,
\ee
(neglecting terms proportional to the up and down quark masses).

Some care must be taken in computing the tree level component of the amplitudes proportional to
$\G_0$, since the related chiral operator (the first in \eq{l.chiral}) allows for a non-vanishing
$K\to0$ transition that must be rotated away, in agreement with the FKW theorem \cite{FKW}.

\medskip

At the one-loop level in chiral perturbation theory, vertex and wave function renormalizations due
to chiral loops appear, which are displayed in \fig{renvertex} and \fig{renwf}, respectively.  We
evaluate the chiral loops in dimensional regularization, and subtract the divergences according to
the $\overline{\rm MS}$, consistently with the \xQM determination of the chiral coefficients.

\pagebreak[3]

At order $O(p^2)$ only the chiral loop corrections to the $\G_0$ term need to be included.  The
resulting analytical expressions for the amplitudes are complicated polynomial and logarithmic
functions of the meson masses. We find it useful to report the following semi-numerical forms of the
isospin amplitudes
\bea
A_0^{\chi_1} &=& \frac{4}{\sqrt{3}}  \frac{\G_0f_\pi^2}{f^5}
\left[1.36+0.46 \,i +0.46 \ln\mu^2\right],
\label{Ax1} \\
A_2^{\chi_1} &=&  \sqrt{\frac{2}{3}}\frac{\G_0f_\pi^2}{f^5}
\left[0.20 + 0.20 \,i + 0.051  \ln\mu^2 \right],
\label{Ax2}
\eea
where $\mu$ is in units of GeV, and we have explicitly factored the tree level amplitudes in
front. The numerical coefficients corresponds to the values of the meson parameters given in
\Table{parvalues}.  The absorptive part of the amplitudes stems from the last diagram in
\fig{renvertex}, as it follows from the Cutkosky cuts. We stress though that in order to obtain the
absorptive part at a given order in the perturbative expansion the amplitude must be evaluated at
the next order.

\begin{table}[t]
\renewcommand*{\arraystretch}{1.35}
\begin{tabular}{lcr}
\hline\hline
$f_\pi$ &\hspace*{.54\columnwidth}& $0.092$ GeV \\
$f_K$         && $0.113$ GeV \\
$m_\pi$       &&  $0.137$ GeV  \\
$m_K$         &&   $0.498$ GeV  \\
$m_\eta$       && $0.547$ GeV \\
$\Lambda_\chi$ && 0.82 GeV \\
$f_1$         && $0.087^{+0.012}_{-0.014}$ GeV\\
M             && $0.200^{+.0.005}_{-0.003}$ GeV\\
$\qq$         && $-(0.240^{+0.030}_{-0.010}\ \rm GeV)^3$ \\
\hline\hline
\end{tabular}
\caption{Values of the physical parameters and phenomenological ranges of the non-perturbative
  \xQM\ parameters used in the numerical analysis.}
\label{parvalues}
\end{table}

These expressions allow us to appreciate the impact of the chiral loops on the full amplitudes.
While they are small in the isospin-two amplitude, the isospin-zero projection receives a sizable
chiral loop renormalization, that is responsible for most of its deviation from the Vacuum
Saturation Approximation (VSA), as we shall discuss next.

It is worth noting that the chiral corrections to the $A_2$ amplitude, \eq{Ax2}, coincide
numerically with those computed for the operators $\Delta Q_{8,7}$ in Ref.~\cite{Bertolini:1997nf},
as mentioned at the end of the previous section. This is due to the fact that the $\Delta Q_{8,7}$
and $Q^{LR}_{1,2}$ share the same $\Delta I = 3/2$ component~\cite{Chen:2008kt}.

\subsection{The $B$-parameters}\label{Bparameters}

\begin{figure*}
\centerline{\includegraphics[width=2\columnwidth]{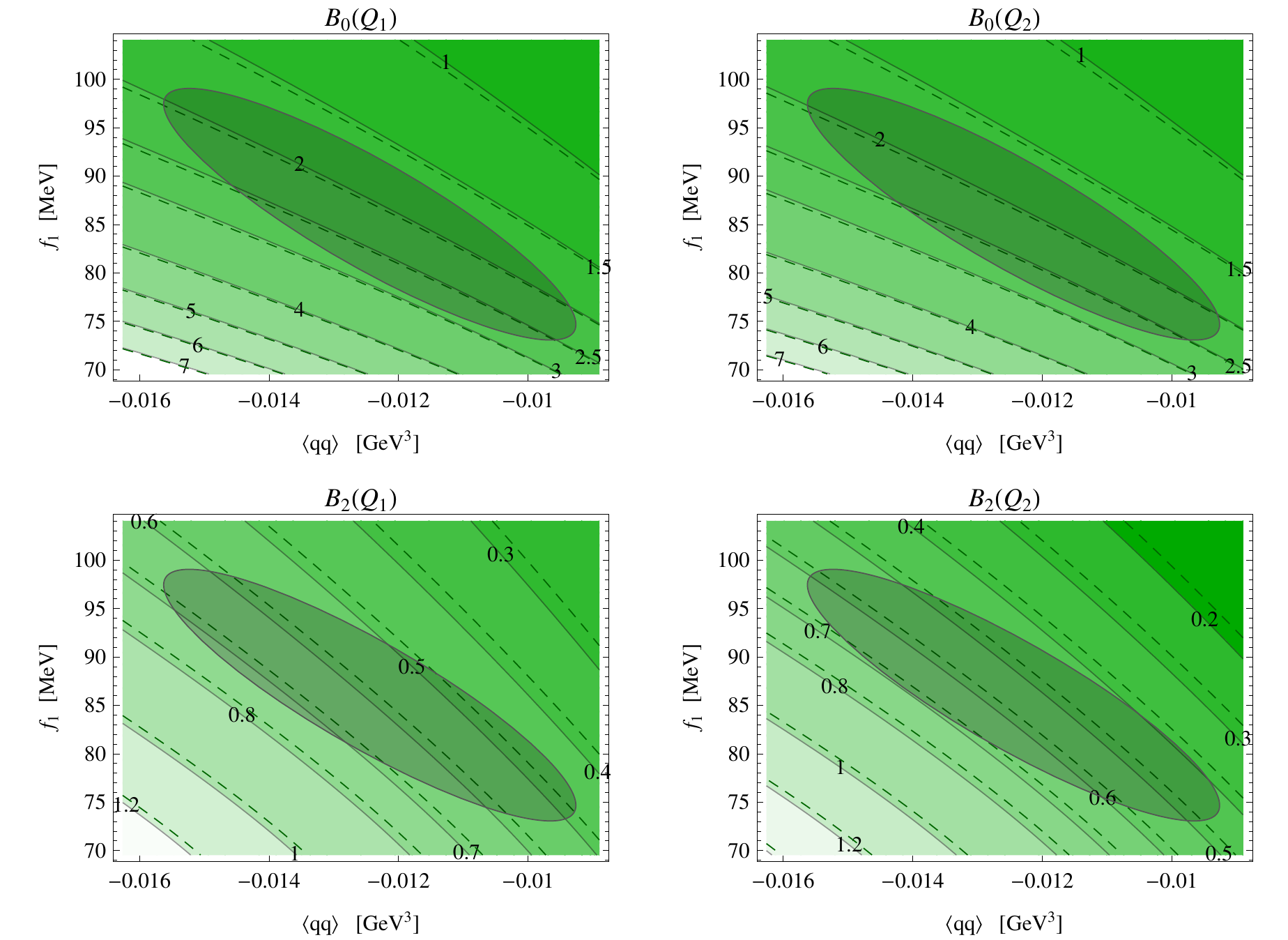}}
\caption{Contour levels of $B_0$ and $B_2$ for $Q_1^{LR,RL}$ (left) and $Q_2^{LR,RL}$ (right panels) in
  the HV and NDR renormalization schemes (continuous and dashed contours respectively). The ellipse
  marks the correlated range of the parameters $\qq$ and $f_1$ in the phenomenological fit of the
  \xQM, as discussed in the text.}
\label{Q12ell}
\end{figure*}

\noindent
A convenient way to show the results is to normalize the $K\to \pi\pi$ matrix elements to their VSA
values.  By denoting $\langle Q\rangle_{0,2}=\langle \pi\pi, I=0,2|Q|K\rangle$, where the subscripts
refer to the isospin components, the $B$-parameters are defined as:
\be
\label{defB}
B_{0,2}\equiv \frac{{\rm Re}\, \langle Q\rangle_{0,2}^{\text{model}}}{\langle Q\rangle_{0,2}^{\text{VSA}}}\,.
\ee

The reference VSA values for the $Q_1^{LR}$ operator can be written as~\cite{Ecker:1985vv}
\bea
\langle Q_1^{LR}\rangle_0^{\text{VSA}} &=&\frac{\sqrt{2} (X+9 Y+3 Z)}{3 \sqrt{3}}\,,  \\
\langle Q_1^{LR}\rangle_2^{\text{VSA}}&=& \frac{1}{3} \sqrt{\frac{1}{3}} (X-6 Z)\,,
\eea
and similarly for $Q_2^{LR}$
\bea
\langle Q_2^{LR}\rangle_0^{\text{VSA}} &=&\frac{\sqrt{2} (3 X+3 Y+Z)}{3 \sqrt{3}}\,,  \\
\langle Q_2^{LR}\rangle_2^{\text{VSA}} &=&\frac{1}{3} \sqrt{\frac{1}{3}} (3 X-2 Z)\,,
\label{VSA}
\eea
where $X \equiv i\sqrt{2}f_{\pi}(m_{K}^{2}-m_{\pi}^{2})$, $Y \equiv i\sqrt{2}f_{K}A^2$, $Z \equiv
i\sqrt{2}f_{\pi}A^2$ respectively and $A\equiv m_{K}^{2}/(m_{s}+m_{d})$.

By taking the scale $\mu$ at the the chiral perturbation theory cutoff $\Lambda_\chi \simeq
0.82\,$GeV (where the values for the non-perturbative parameters $M$ and $\qq$ in \Table{parvalues}
were obtained in~\cite{Bertolini:1997ir}) and spanning over the model parameter space, we obtain the
values for the different $B_{0,2}(Q^{LR}_{1,2})$.  In \fig{Q12ell} the contour levels of $B_0$ and
$B_2$ as a function of the relevant \xQM parameters are displayed.  Both HV and NDR schemes results
are shown. As one can see, the $\gamma_5$-scheme dependence is quite limited since it appears at
$O(M^2/\Lambda_\chi^2$).

The numerical summary of $B_{0,2}$ for the two LR operators is reported in \Table{Bvalues}
(the same values hold for the RL related operators).
Their uncertainties are evaluated by considering the variation of the relevant parameters, namely
$M$, $\qq$, $f_1$ and $m_s$.  The parameters $f_1$ and $\qq$ have a correlated variation range,
displayed as the shaded ellipses in \fig{Q12ell}. The correlation stems from the dependence on $f_1$
and $\qq$ on the NLO low energy constant $L_5$ in the strong chiral lagrangian as computed in the
\xQM (see App. B in Ref. \cite{Bertolini:1997ir}), namely
\be
L_5 =
-\frac{f_1^4}{8M\qq}\left(1-6\frac{M^2}{\Lambda_\chi^2}\right),
\label{L5}
\ee
by taking into account the present uncertainty on the knowledge of $L_5$ (about $10\%$).  This
correlated variation drives most part of the final uncertainty in $B_0$, $B_2$, the effect of the
constituent quark mass $M$ and of its correlation being numerically irrelevant.  We have
consistently (and conservatively) used for the strange quark mass its PCAC value $m_s(\mu) =
-f_\pi^2 m_K^2/\qq(\mu)$, which for the range of the quark condensate given in~\Table{parvalues}
leads to $m_s(\Lambda_\chi)= 152^{+20}_{-45}$\,MeV.  This is well consistent, albeit with a larger
range, with $m_s(2\,\text{GeV})= 95\pm 5$\,MeV from lattice determinations~\cite{PDG}.

\begin{table}[t]
$
\begin{array}{lcccc}
\hline\hline
 &\hspace*{.18\columnwidth}& \rm NDR &\hspace*{.20\columnwidth}& \rm HV\\[.5ex]
 \hline
B_0(Q_1^{LR,RL})&& 2.00_{-0.39}^{+0.87} && 2.04_{-0.40}^{+0.85}\\[1ex]
B_0(Q_2^{LR,RL})&& 1.95_{-0.37}^{+0.82} && 1.99_{-0.38}^{+0.80}\\[1ex]
B_2(Q_1^{LR,RL})&& 0.64_{-0.17}^{+0.11} && 0.62_{-0.17}^{+0.11}\\[1ex]
B_2(Q_2^{LR,RL})&& 0.59_{-0.18}^{+0.14} && 0.57_{-0.18}^{+0.13}\\[1ex]
\hline\hline
\end{array}
$
\caption{Values of $B_0$ and $B_2$ at $\mu=0.82\,$GeV for $\bra{\pi\pi}Q_{1,2}^{LR,RL}\ket{K}$, in the NDR and HV $\gamma_5$-schemes.}
\label{Bvalues}
\end{table}
\begin{table}[t]
$
\begin{array}{lcccc}
\hline\hline
 &\hspace*{.18\columnwidth}& \rm NDR &\hspace*{.20\columnwidth}& \rm HV\\[.5ex]
 \hline
B_0(Q_1^{LR,RL})&& 1.84_{-0.36}^{+0.85} &&  1.87_{-0.37}^{+0.84} \\[1ex]
B_0(Q_2^{LR,RL})&& 1.82_{-0.35}^{+0.82} &&  1.84_{-0.36}^{+0.81} \\[1ex]
B_2(Q_1^{LR,RL})&& 0.55_{-0.15}^{+0.09} && 0.54_{-0.15}^{+0.09}\\[1ex]
B_2(Q_2^{LR,RL})&& 0.52_{-0.15}^{+0.10} && 0.51_{-0.15}^{+0.10}\\[1ex]
\hline\hline
\end{array}
$
\caption{Same as \Table{Bvalues} at  $\mu=2\,$GeV.}
\label{Bvalues2}
\end{table}

The $B$-parameters for the LR current-current operators $Q_{1,2}^{LR}$ turn out to have a size
comparable to that of the standard electroweak penguins $Q_{7,8}$~\cite{Bertolini:1998vd}. For the
isospin two component of the amplitudes one expects, on the basis of the symmetry arguments
discussed earlier, the parameters $B_2(Q_{1,2}^{LR})$ to be the same as the corresponding
$B_2(Q_{8,7})$. The numerical difference is due to the $O(p^4)$ corrections included in
Refs.~\cite{Bertolini:1997ir,Bertolini:1997nf} and here neglected.

For future reference we also give in \Table{Bvalues2} the values of $B_{0}$, $B_{2}$ at the
renormalization scale $\mu=2\,$GeV, calculated by taking into account the anomalous dimension matrix
of the $Q_{1,2}^{LR}$ operators~\cite{Cho:1993zb} as well as the running of $\qq$ and $m_s$ (related
by PCAC). As one sees, within the uncertainties, the values of $B_0$ and $B_2$ can be considered
scale independent between $\Lambda_\chi$ and 2\,GeV, in agreement with the leading role of the
$O(p^0)$ coefficient $\G_0$ and its dependence on $\qq\!{}^2$, analogous to the VSA.


\section{Conclusions}

\noindent
In this work we considered the calculation of hadronic matrix elements of the $\Delta S=1$
left-right four-quark operators $Q_{1,2}^{LR}$ which are present in popular extension of the SM.
Because of their possible tree level origin, they are potentially the source of large contributions
of New Physics to kaon hadronic decays, thus giving rise to stringent constraints on the new-physics
scales and/or couplings.  A paradigmatic example is the LR symmetric model, where the above
operators are generated at a large scale by $W_L-W_R$ mixing.  The possibility of a TeV size
right-handed scale, together with the absence of loop suppression in the Wilson coefficients, may be
the source of sizable contributions of LR current-current operators to direct CP violation in the
kaon sector.  The $Q_{1,2}^{LR,RL}$ operators are present in minimal extensions of the SM Yukawa
sector and in SUSY extensions as well.

Among the complete set of $\Delta S=1$ four-quarks operators, these were the only one for which an
evaluation of the relevant $\langle \pi\pi|Q_{1,2}^{LR}|K\rangle$ matrix elements was missing.  We
addressed the calculation of these hadronic matrix elements in the context of the Chiral Quark
Model.  To this aim, the complete $O(p^2)$ $\Delta S=1$ effective chiral lagrangian was constructed.
This allowed us to perform a complete evaluation of the $K\to \pi\pi$ matrix elements at $O(p^2)$,
which includes the one-loop chiral contributions. The computation was performed in both NDR and HV
$\gamma_5$ renormalization schemes.  The $K\to \pi\pi$ amplitudes for $Q_{1,2}^{LR}$ were found to
be similar to those of the standard penguin operators $Q_{7,8}$ respectively (which are partially
related to the LR current-current ones by symmetry arguments). We compared our results, obtained at
the chiral breaking scale, with those of the simple factorization (VSA), showing deviations within
$50\%$. For a convenient comparison with forthcoming (and hopefully ultimate) lattice results, the
values of the matrix elements are also shown at the scale of 2\,GeV.


\section*{Acknowledgments}

\noindent
S.B. acknowledges partial support by the italian MIUR grant no.~2010YJ2NYW\_001 and by the EU Marie
Curie ITN UNILHC grant no.~PITN-GA-2009-237920.  A.M. acknowledges SISSA for hospitality during the
conclusion of this work.

\def\arxiv#1[#2]{\href{http:/arxiv.org/abs/#1}{[#2]}}

\end{document}